\documentclass[twocolumn,prc,aps,nofootinbib]{revtex4}

\usepackage[latin9]{inputenc}
\usepackage{color}
\usepackage{amsmath}
\usepackage{amsthm}
\usepackage{amssymb}
\usepackage{graphicx}

\usepackage{ulem}

\makeatletter

\usepackage{epsfig}
\usepackage{amsthm}
\usepackage{tikz}
\usetikzlibrary{quantikz}

\usepackage{bm}

\usepackage{color}

\makeatother

\begin{document}
\title{Accessing ground state and excited states energies in many-body system after symmetry restoration using quantum computers} 
\author{Edgar Andres Ruiz Guzman }
\email{ruiz-guzman@ijclab.in2p3.fr}

\affiliation{Universit\'e Paris-Saclay, CNRS/IN2P3, IJCLab, 91405 Orsay, France}
\author{Denis Lacroix }
\email{denis.lacroix@ijclab.in2p3.fr}

\affiliation{Universit\'e Paris-Saclay, CNRS/IN2P3, IJCLab, 91405 Orsay, France}
\date{\today}

\begin{abstract}
We explore the possibility to perform symmetry restoration with the variation after projection technique 
on a quantum computer followed by additional post-processing. The final goal is to develop configuration interaction techniques based on
many-body trial states pre-optimized on a quantum computer.  We show how the projection method used for symmetry restoration can prepare optimized states 
that could then be employed as initial states for quantum or hybrid quantum-classical algorithms. We use the quantum phase estimation 
and quantum Krylov approaches for the post-processing.  The latter method combined with the quantum variation after projection (Q-VAP)
 leads to very fast convergence towards the ground-state energy. The possibility to access excited states energies is also discussed. Illustrations of the 
 different techniques are made using the pairing hamiltonian.   
 \end{abstract}
\keywords{quantum computing, quantum algorithms}

\maketitle 

\section{Introduction}

The development of novel generations of nuclear many-body forces has 
promoted {\it ab-initio} methods as a tool of choice to describe 
microscopically atomic nuclei from the underlying bare nucleon-nucleon interaction. 
Nowadays, a large variety of approaches are being developed that, depending 
on the underlying approximations, can be applied to certain
regions of the nuclear chart. 
Among the  important challenges that should be considered for the applicability of {\it ab-initio} theories, we mention 
the following two. Firstly, methods treating exactly the many-body systems face the problem of the exponential growth 
of the Hilbert space size when the number of single-particle states increases. This is for instance the case of the Faddeev--Yakubovski \cite{Fri93, Glo93,Nog97}, 
Green's function Monte Carlo \cite{Pud97,Wir98,Wir00} and No-Core Shell Model (NCSM)  \cite{Qua08, Nav09} approaches that are restricted to rather light systems.   

In view of this first difficulty, several approaches have been proposed in the last 20 years that have a more appropriate scaling (polynomial 
scaling generally) to tackle up to medium-mass nuclei. Among these methods, we mention the Many-Body Perturbation Theory (MBPT) \cite{Isa09,Tic16,Hu16}, Coupled Cluster (CC) \cite{Bar07}, 
In--Medium Similarity Renormalization Group (IMSRG) \cite{Her16} or Self-Consistent Green Function (SCGF) \cite{Dic04} methods. 

Another important challenge that might be  considered in atomic nuclei, especially for the precise description of open shell nuclei or medium/heavy systems, is the possibility to take advantage of the symmetry breaking techniques followed by symmetry restoration \cite{Rin80,Bla86,Ben03, Rob18,She19}. 
An intensive effort is now being made to extend some of the approaches listed above in such a way that they start from a symmetry-breaking trial state: the Bogoliubov Many-Body Perturbation Theory (BMBPT) \cite{Tic18,Dem21}, the Gorkov Self-Consistent Green Function (GSCGF) 
\cite{Som11,Som14,Som20} and the Bogoliubov Coupled Cluster (BCC)  \cite{Sig15, Dug17}. We note that these techniques have been sometimes supplemented by symmetry restoration through projection techniques eventually followed by further configuration-interaction (CI) diagonalization in a reduced Hilbert space \cite{Rip17,Rip18,Fro21}.  Among them, we mention the  Projected Bogolyubov MBPT \cite{Lac12}, the Projected Bogoliubov Coupled Cluster (PBCC) \cite{Qiu17,Qiu19}, the Projected QRPA \cite{Gam12} or very recently the projected generator coordinate method - Perturbation Theory (PGCM-PT) \cite{Fro21a, Fro21b, Fro21c}. Still, at present, the later methods have been mainly tested in rather simple models. For a comprehensive recent review, we recommend the reference \cite{Fro21}.

In view of the current scientific emulation, we explore here the possibility to follow a strategy of symmetry breaking--symmetry restoration
followed eventually by further post-processing using quantum computers. We believe that such exploration is particularly timely with the current boost in building quantum devices. 
The current status, called NISQ (Noisy-Quantum Intermediate Quantum) period does not allow for performing complicated 
many-body calculations. Nevertheless, an increasing number of pilots applications are being nowadays made in different fields of physics 
 \cite{Lan10,Bab15,OMa16,Col18,Hem18,Mac18,Dum18,Lu19,Rog19,Du20,Klc18,Klc19,Ale19,Lam19}. 
 
We consider here a long-term strategy to prepare future applications beyond the NISQ period. We discuss first below a method to prepare 
a many-body trial state on a quantum computer that takes advantage of the symmetry breaking--symmetry restoration technique. 
A first milestone in that direction is achieved by optimizing a parametric state using the standard Variational Quantum Eigensolver (VQE) technique \cite{Mcc17,Fan19,Cao19,McA20,Bau20,Bha21} 
leading to a method we call hereafter Quantum--Variation After Projection (Q-VAP). 
We then explore the possibility to use this state for further post-processing either directly 
on the quantum computer or using hybrid quantum--classical technologies.       
  
\section{Variation After Projection on a quantum computer}


The strategy to perform the Variation After Projection on a quantum computer follows closely the 
method that is used in classical computers. We consider a quantum many--body 
system that is mapped onto a set of $N$ qubits labelled by $i=0, N-1$.  A complete basis of the system
is then given by the states $ \{ \bigotimes_{i=0}^{N-1} | s_i \rangle \} $ where $| s_i =0_i ,1_i \rangle$
correspond to the two states associated to the $i^{th}$ qubit. A wave--function can be written in the full Fock
space as:
\begin{eqnarray}
| \Psi \rangle = \sum_{s_i\in\{0,1\}} \Psi_{s_1, \cdots, s_N} | s_1, \cdots, s_n \rangle. \label{eq:generalphi}
\end{eqnarray}
Our first objective here is to obtain wave functions in a quantum computer that can properly describe 
interacting fermions under the action of a many--body Hamiltonian $H$.   One of the strategies used nowadays to approach this problem on a quantum computer 
is to
express the trial state vector in terms of a set of parameters 
denoted as $\{ \theta_i\}_{i=1, \dots, N_{\theta}}$. For recent reviews on the subject see for instance \cite{Mcc17,Fan19,Cao19,McA20,Bau20,Bha21}.

In general, the trial state that is optimized during the minimization of the energy is obtained from a set of unitary operations starting from 
the vacuum, denoted hereafter simply as $| 0 \rangle \equiv  \bigotimes_{i=0}^{N-1} | 0_i \rangle $, such that:
\begin{eqnarray}
| \Psi (\{ \theta_i \} ) \rangle = \prod_{k=1}^{N_\theta} U_k (\theta_k) | 0 \rangle.  \label{eq:trial}
\end{eqnarray} 
By using the expectation of the Hamiltonian as the cost function, variational methods are firstly targeted to reproduce the 
ground state energy of the problem. The precision on the energy will obviously intimately depend on the transformations 
that are used in Eq. (\ref{eq:trial}). One of the issue for quantum computers is the possibility to reduce the circuit 
depth by using the symmetry of the underlying problem (see for instance \cite{Mol16, Lui19,Gar20}).
  
Here, we take a different point of view and suppose that the trial state defined by Eq. (\ref{eq:trial}) might breaks some of the symmetries
of the underlying Hamiltonian. This technique where symmetry-breaking (SB) states is used is rather standard in many fields of physics \cite{Rin80,Bla86} and is known as a very accurate method to grasp complex internal correlations when the system encounters spontaneous symmetry breaking.  Typical examples are superfluid systems where the $U(1)$ symmetry associated to the particle number conservation is broken by forming 
Cooper pairs. As underlined in the introduction, atomic nuclei are such complex systems where it can be advantageous to break symmetries like 
particle number, parity or rotational symmetry. The possibility to use  SB states in quantum computers has already been promoted for instance in Refs. \cite{Ver09,Jia18,Lac20,Kha21}. 

 One pre-requisite to obtain precise meaningful description of the ground state 
energy is that the symmetries that are initial broken are restored in a second step. Symmetry restoration (SR) by projection is nowadays a standard tool in atomic nuclei. One usually 
distinguishes the projection after variation (PAV) and variation after projection (VAP)\cite{Ben03,Rob18,She19}. 
Let us assume that a certain symmetry $S$ is broken by the state (\ref{eq:trial}) and denote generically 
by ${\cal P}_{S}$ the projector associated to the restoration of this symmetry. In the PAV approach, the expectation value of the 
energy of the SB state given by 
\begin{eqnarray}
E_{\rm SB }(\{ \theta_i \}) &=& \langle \Psi (\{ \theta_i \} ) | H | \Psi (\{ \theta_i \} ) \rangle,  \label{eq:esb}
\end{eqnarray}
is minimized.  Then, the PAV energy is  directly given by the expectation value of the Hamiltonian after projection of the trial state:
\begin{eqnarray}
E_{\rm PAV} (\{ \theta_i \}) &\equiv \displaystyle \frac{\langle \Psi (\{ \theta_i \} ) | H {\cal P}_{S} | \Psi (\{ \theta_i \} ) \rangle }{ \langle \Psi (\{ \theta_i \} ) | {\cal P}_{S} | \Psi (\{ \theta_i \} ) \rangle } \label{eq:pav}
\end{eqnarray} 
where we use the fact that $ {\cal P}^2_{S} =0$ and that $[H, {\cal P}_{S}] = 0$. 

The VAP approximation is more challenging and consists in minimizing directly the energy given by Eq. (\ref{eq:pav}) for the projected state. This energy is denoted by $E_{\rm VAP} (\{ \theta_i \})$. Thanks to the use of the variational principle and because both PAV and VAP 
states belongs to the same Hilbert subspace that respect the restored symmetry, we automatically have the property $E_{\rm GS} \le E_{\rm VAP} \le E_{\rm PAV }$ at the minimum of the VAP method. We denoted by $E_{\rm GS}$ the ground state energy.   

A first milestone in transposing the SB-SR methodology on quantum computers has been reached in Ref. \cite{Lac20} where a quantum algorithm was proposed to perform symmetry restoration. In this reference, the Quantum-Phase-Estimation (QPE) algorithm 
\cite{Fan19, Nie02,Hid19,Ovr03,Ovr07} was used to perform the projection. The QPE method is originally designed to obtain the eigenvalues 
and eigenvectors of a unitary operator using the Quantum Fourier Transform (QFT) \cite{Nie02} together with a set of additional ancillary qubits. 
Repeated measurements of the ancillary qubits give access to the different eigenvalues. Provided that the number of qubits is sufficient to separate each eigenvalue, 
the state after each measurement is projected onto the set of eigenvectors associated to this eigenvalue.  

The original idea that was proposed
in Ref. \cite{Lac20} is that the QPE method can be directly used as a projector for symmetry restoration.  For this, it is sufficient to use the QPE 
with an operator with known eigenvalues such that each eigenvalue is associated with subspaces of the total Fock space having the proper symmetry.  An illustration was given in Ref. \cite{Lac20} where the $U(1)$ symmetry was restored using an operator proportional to the 
particle number. Another example was given in \cite{Siw21} where the method was applied to project spin states onto eigenstates of the total spin ${\bf S}^2$.    

Up to know, the technique proposed in \cite{Lac20} has only been used for the SR and, as far as we know, has never been combined with a variational method on a quantum computer.  Below, we give an example of use of the projection technique used together variational quantum methods. We then perform the equivalent of the PAV and VAP methods on a quantum computer. In analogy with their counterparts in classical computers, we call the two procedures Quantum-PAV (Q-PAV) and Quantum-VAP (Q-VAP) respectively. 
In the following, as the first step of our study, we consider a schematic illustration of the methods Q- PAV and Q-VAP. We mention that projected states were used  in \cite{Kha21}, employing a completely different projection technique that was supplemented by an additional correlator (the so-called pair-hopper operator).

\subsection{Application to the pairing Hamiltonian}

As an illustration of application of the Q-PAV and Q-VAP methodology, we consider here a pairing Hamiltonian \cite{Von01,Zel03,Duk04,Bri05}.
The system is composed of fermions distributed on a set of doubly--degenerated single-particle levels $p=0,N-1$. The two-body 
Hamiltonian of the system is written in second quantized form as 
\begin{eqnarray}
H & = &\sum_p\varepsilon_p \hat N_p - g \sum_{pq} \hat P^\dagger_p \hat P_q,   \label{eq:hampairing}
\end{eqnarray}
where the operator entering in the Hamiltonian are respectively the pair occupations and pair creations operators defined as:
\begin{eqnarray}
\hat{N}_{p} & = & a_{p}^{\dagger}a_{p}+a_{\bar{p}}^{\dagger}a_{\bar{p}}, ~~\hat{P}_{p}^{\dagger}  =  a_{p}^{\dagger}a_{\bar{p}}^{\dagger}.
\end{eqnarray}
 $(a_{p}^{\dagger},a_{\bar{p}}^{\dagger})$ are creation operators of time-reversed single-particle states associated to the energies $\varepsilon_p$. 
  
This hamiltonian, that gives a schematic description of superfluid systems, has already been used as a test-bench for quantum 
computers algorithms  using different fermions to qubits mappings \cite{Ovr03,Ovr07,Lac20,Kha21,Rui21}. The mapping can be made using 
the standard Jordan-Wigner transformation (JWT) \cite{Jor28,Lie61,Som02,See12,Dum18,Fan19} either at the level of the single-particle states \cite{Ovr03,Ovr07,Lac20} or directly at the level of the pair creation operators \cite{Kha21,Rui21}. We consider here the second method that has the advantage to reduce by a factor $2$ the number of qubits to encode the problem and the shortcoming that only even systems can be considered.  For each qubit $p$, we introduce the standard Pauli matrices denoted by  $(X_p, Y_p, Z_p)$ that are completed by the identity $I_p$. Mapping directly the pairs using the JWT method, we have the correspondence:
\begin{eqnarray}
\hat P^\dagger_p &\longrightarrow& P^+_p= \frac{1}{2} (X_p - i Y_p) = \left[ \begin{array}{rr}0&0\\ 1 &0\end{array}\right]_p,  \nonumber \\
\hat N_p &\longrightarrow& N_p = 1 - Z_p= \left[ \begin{array}{rr}0&0\\ 0 &2 \end{array}\right]_p .   \nonumber 
\end{eqnarray}
The JWT mapping gives the equivalent Hamiltonian acting on the $N$ qubits:  
\begin{eqnarray}
H & = & \sum_{p} (\varepsilon_{p} - g/2)\left[1-Z_{p}\right]-\frac{g}{2}\sum_{p>q} \left[X_{p}X_{q}+Y_{p}Y_{q}\right].
\end{eqnarray}  

The pairing problem is an archetype of a problem where it is advantageous to break a symmetry in order to treat certain correlations. 
In this model, above a certain threshold of the two-body interaction strength, 
the system encounters a  transition from a normal to a superfluid phase. Then, the problem becomes highly non-perturbative. The internal correlations can then be treated while maintaining relatively simple trial states, provided that the $U(1)$ symmetry associated with the particle number is broken. This is actually the essence of the BCS and HFB theory \cite{Rin80,Bri05}. 

\begin{figure}[htbp] 
\includegraphics[scale=0.3,angle=-90]{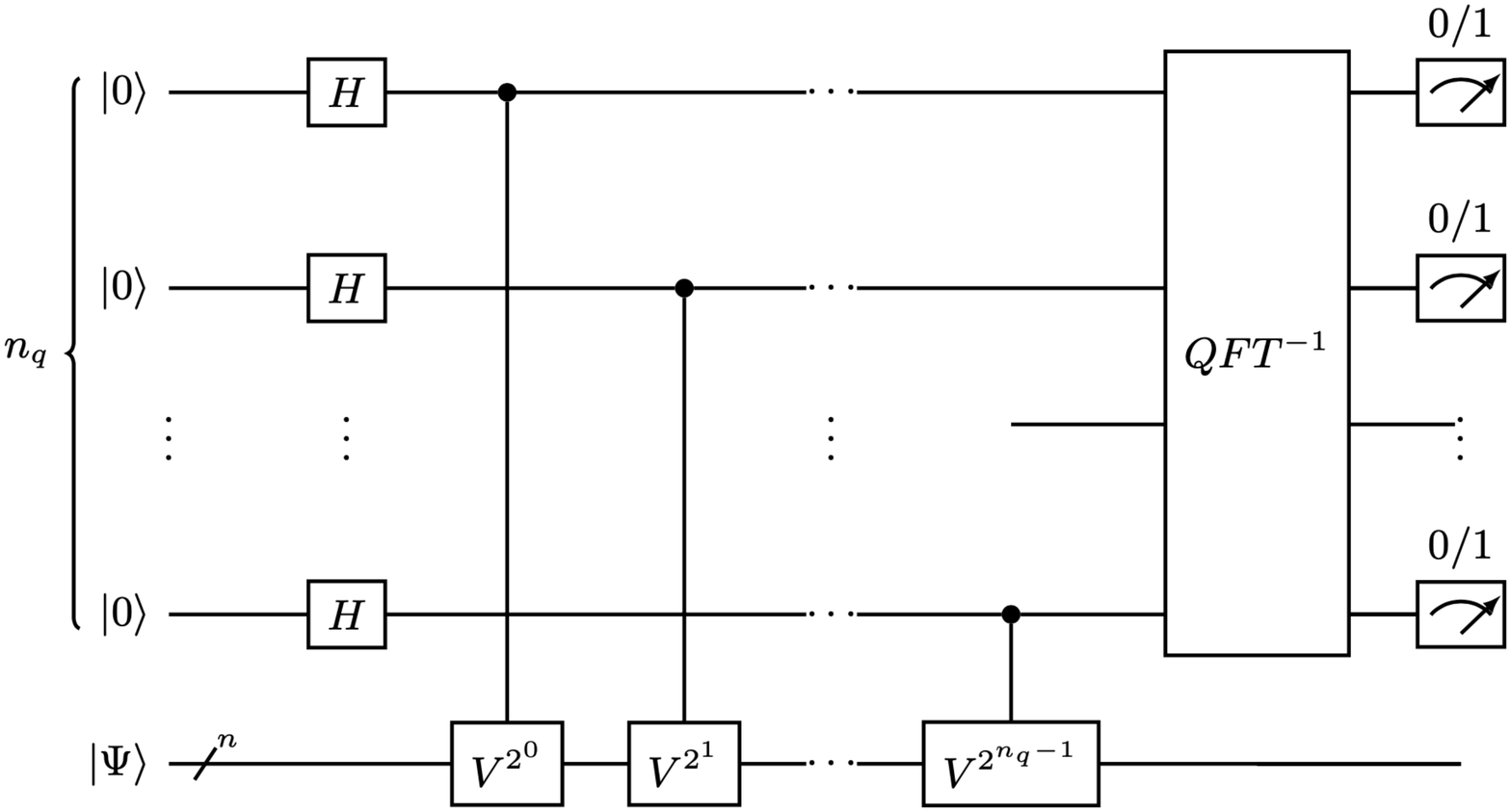} 
    \caption{Schematic view of the QPE method applied to the operator $V$ with $n_q$ ancillary qubits. The circuit shown here and the ones in the following are made using
the quantikz package  from Ref. \cite{Kay18}.  }
    \label{fig:qpe}
\end{figure}

\subsubsection{Quantum BCS ansatz}

As a starting wave-function, we consider the standard BCS ansatz, that, with our method of directly encoding the pairs and using the convention of Ref. \cite{Bri05},  takes the form:
\begin{eqnarray}
| \Psi(\{ \theta_p \}) \rangle &=&  \bigotimes_{p=1}^{N-1} \left[ \sin(\theta_p) | 0_p \rangle + \cos(\theta_p) |1_p \rangle \right] \label{eq:bcsansatz}\\
&=& \prod_{i=0}^{N-1} R^p_Y (\pi - 2\theta_p)  \bigotimes_{p=0}^{N-1}  |0_p \rangle  \nonumber 
\end{eqnarray}    
with the convention $R^p_Y(\varphi) = e^{-i Y_p \varphi/2}$. 
Given that the quantum circuit used to prepare this state corresponds to independent rotations of each qubit.

\subsubsection{Particle number projection}
\label{eq:proj}

The state (\ref{eq:bcsansatz}) mixes different particle numbers. Here, we follow \cite{Lac20,Siw21} and use the QPE to project the BCS state on a given number of particles. The QPE is rather well documented  \cite{Nie02,Fan19} and we only give here the useful ingredients for the following discussion.

Assuming a unitary operator $V$ with a set of eigenvalues written as $e^{2\pi i \varphi_\alpha }$ and associated with the eigenstates $| \varphi_\alpha \rangle$, the QPE is a practical way to obtain the phases $\{ \varphi_\alpha \}$ and the eigenstates with some precision, assuming that all $\varphi_\alpha$ verifies:
\begin{eqnarray}
0 \le\varphi_\alpha < 1. \label{eq:constphi}
\end{eqnarray} 
 The QPE method works as follows. The method uses a set of ancillary qubits $n_q$.  A set of controlled-$V^{2^j}$ operations, with $j=0, n_q-1$, is performed to transfer the information about the eigenstates of $V$ to the ancillary qubits. The associated circuit is shown in Fig. \ref{fig:qpe}. The approximate values of $\varphi_\alpha$ and the projection onto the associated eigenstates are obtained from the measurement of the ancillary qubits after performing an inverse Quantum Fourier Transform on the $n_q$ quantum register.  In practice, the 
projection is performed by appropriately selecting the operator $V$ and the number of qubits. An illustration of operator for the 
particle number projection was given in \cite{Lac20}. Here, we consider the following operator: 
\begin{eqnarray}
V = \exp\left( 2 \pi i \frac{N_{\rm P}}{2^{n_q}} \right) = \prod_{p=0}^{N-1} \left(
\begin{array}{cc}
 1     &  0  \\
 0     &  e^{i \pi/2^{n_q-1} } 
\end{array} 
\right)_p ,  \label{eq:par_nb_op}
\end{eqnarray}       
where we use the operator $N_{\rm P} = \sum_p N_p /2$ that counts the number of pairs. The operator $N_{\rm P}$ has eigenvalues $0, \cdots , N$ for $N$ 
levels. We recognize on the right side of (\ref{eq:par_nb_op}) a simple product of phase operators. 
 The condition $\varphi_\alpha < 1$ fixes the minimal number of ancillary qubits to be used for properly resolving the different 
eigenvalues. This gives the constraint: 
\begin{eqnarray}
n_q >  \frac{\ln N}{\ln 2} .\label{eq:constraint}
\end{eqnarray}

In practice, the method we propose for the projection works like a filter for the SB initial states. After each measurement of the 
ancillary qubits, we obtain a binary number $\delta_1 \cdots \delta_{n_q}$ that corresponds to the binary fraction of one of the eigenvalues 
$\varphi_\alpha$, or equivalently, to a given number of pairs denoted as $A_P$. 
After the measurement of the ancillary qubits, the BCS state is projected 
onto the corresponding symmetry restored state with exactly $A_P$ pairs. It is worth noting that, event-by-event, different values of $A_P$ 
can be obtained depending on the initial mixing. The only way to influence the result of the measurement is through the initial mixing in the BCS state. Most often, we are interested in a precise value of $A_P$ as the outcome of the quantum projection procedure. This implies that part of the events are rejected after measurements and only 
events with the targeted value of $A_P$ are retained for further post-processing.   

\subsubsection{Illustration of Q-PAV and Q-VAP for the pairing Hamiltonian}

\begin{figure}
\begin{centering}
\includegraphics[scale=0.55]{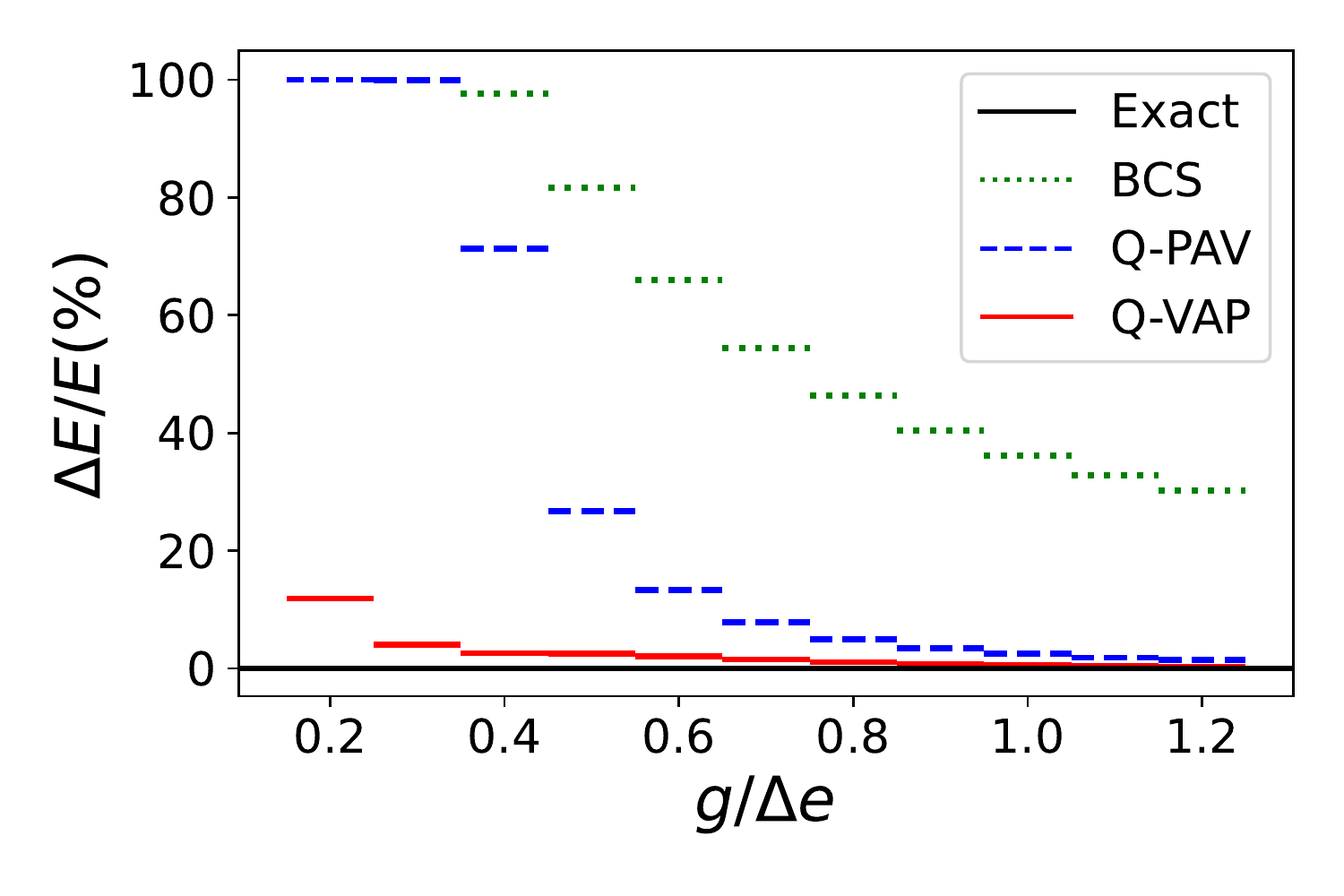} 
\par\end{centering}
\caption{Illustration of the precision in energy (using the quantity defined by Eq. (\ref{eq:ecpercent}))
obtained using the BCS (green dotted line), Q-PAV (blue dashed line) and Q-VAP (red solid line) for the pairing problem with $8$ particles on $N=8$ 
equidistant levels for  $g/\Delta e$ ranging from $0.2$ until $1.2$ with a $0.1$ step. The black solid line indicates the exact result, i.e.,  $\Delta E/E = 0$. 
Results have been obtained using the Hybrid Quantum-Classical minimization procedure 
and projection procedure described in the text. }
\label{fig:figbcsP}
\end{figure}

We show in Fig. \ref{fig:figbcsP} the results obtained using the BCS ansatz for $8$ particles, i.e., 4 pairs,
on $N=8$ doubly degenerated levels and for various interaction strengths. In this figure, we use the correlation energy $E_c$ defined as the total 
energy minus the reference Hartree-Fock energy defined as the energy  of the system when filling the $N/2$ least energetic doubly degenerated levels. The error is then defined as \cite{Rip17}:
\begin{eqnarray}
\frac{\Delta E}{E} (\%) &=& \left| \frac{E^{\rm approx}_c - E^{\rm exact}_c}{E^{\rm exact}_c} \right| \times 100 .\label{eq:ecpercent}
\end{eqnarray} 

In different applications, we consider the case of equidistant single-particle levels with 
$\varepsilon_p =p\Delta e$ ($p=1,\dots,N$). We assume $\hbar =1$ and all quantities are shown in $\Delta e$ units. 
The results shown in Fig. \ref{fig:figbcsP} have been obtained using the Qiskit emulator \cite{qiskit}. 
In addition to the BCS result, the hybrid quantum-classical methods Q-PAV and 
Q-VAP were used to obtain the set of $\{ \theta_p \}_{p=1,8}$ that minimizes the energy with an additional constraint 
on particle number. 

In the present implementation, the expectation values of the Hamiltonian are obtained by first decomposing the Hamiltonian as a 
sum of Pauli  chains denoted by $\{ V_l \}$ such that:  
\begin{eqnarray}
H = \sum_l \beta_l V_l. \label{eq:betal}
\end{eqnarray}
 Each Pauli chain $V_l$ is composed of the product of Pauli matrices. 
 Then $\langle H \rangle$ is obtained by computing each expectation value $\langle V_l \rangle$ separately using a standard Hadamard test.

The different steps for the Hybrid Quantum-Classical calculation are closely related to the standard way to solve the BCS on classical computer except that some of the tasks 
are performed by the quantum computer. Explicitly, we use the following iterative procedure (i) some initial values for the set of angles $\{\theta_p \}$ and for the Fermi energy $\lambda$ are chosen. 
(ii) While the condition $|\langle N_P \rangle - A_P | \le {\varepsilon_{\rm tol}}$ is not satisfied (where $\varepsilon_{\rm tol}$ is a tolerance parameter for the difference set manually with a value in our case of $10^{-3}$), the following steps are performed.  
(ii.i) The following cost function is minimized respect to the set of parameters $\{ \theta_p \}$:
\begin{eqnarray}
\mathcal{C}(\{ \theta_p \}) = \langle \Psi(\{ \theta_p \})  | H  - \lambda (N_{\rm P} - A_P) | \Psi(\{ \theta_p \}) \rangle  \label{eq:cost} 
\end{eqnarray} 
where $A_P$ is a constant that is fixed a priori.  In the present case, it is set to $A/2$ where $A$ is the number of particles of interest. The minimization is performed using the COBYLA optimizer.
The expectation value over $H$ is obtained using the decomposition (\ref{eq:betal}) and computing each $\langle \Psi(\{ \theta_p \})  | V_{\rm l} | \Psi(\{ \theta_p \}) \rangle$ on a quantum computer. The expectation over $N_{\rm P} $ is calculated by a classical computer using the formula $\langle \Psi(\{ \theta_p \})  | N_{\rm P} | \Psi(\{ \theta_p \}) \rangle = \sum_p \cos^2(\theta_p) $. (ii.ii) Using the set of optimized parameters $\{ \theta_p \}$, we compute the variation of the Fermi energy $\lambda$. (ii.iii) Lastly, we use the new set of $\{ \theta_p , \lambda \}$ to restart the process at (ii.i).

The different calculations, i.e. BCS, Q-PAV and Q-VAP use the same procedure but differs in the circuits to construct 
the trial states. In the BCS case, the wave-function (\ref{eq:bcsansatz}) is used in the minimization and the cost function is the one shown in (\ref{eq:cost}). The BCS energy shown in Fig. \ref{fig:figbcsP} is computed using the decomposition (\ref{eq:betal}). In the Q-PAV case, the same minimization is performed but after convergence, the BCS state 
is projected onto the pair number $A/2$. The Q-PAV energy reported in Fig. \ref{fig:figbcsP} is computed using the decomposition (\ref{eq:betal}) and the projected state.
The Q-VAP case is more complex because the projected state is directly used in Eq. (\ref{eq:cost}) during the iterative process which means in practice that the QPE 
algorithm is used at each step to project onto a specific particle number before calculating the expectation values. 

The energies obtained in Fig. \ref{fig:figbcsP} using the Hybrid quantum-classical algorithms have been compared to their purely classical counterparts (not shown here). 
In all cases, very good agreements were found validating the combined projection-optimization methodologies. 

\section{Quantum and Hybrid classical-quantum post-processing}

The construction of symmetry-restored states is the first step of a more ambitious goal,
 which is to obtain eigenvalues of a complicated many-body problem.  As a first 
illustration of pure quantum post-processing, we apply the QPE technique
we used above for restoring symmetries, but, this time, to obtain eigenvalues and eigenvectors 
of the many-body Hamiltonian. We then discuss alternative methods that could
reduce the quantum resources.

\begin{figure*}
\begin{centering}
\includegraphics[scale=1.1]{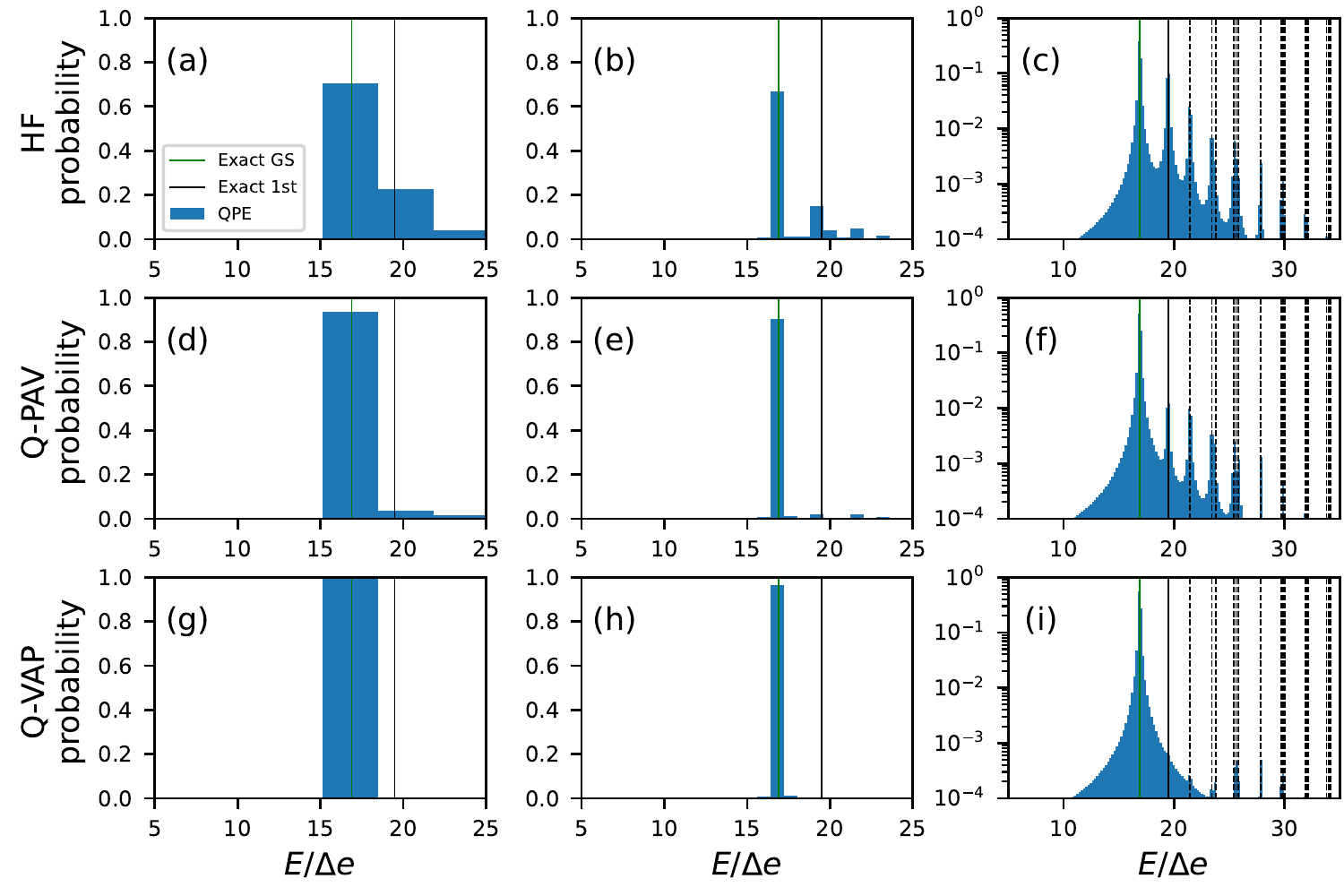}
\par\end{centering}
\caption{Illustration of the results obtained by the QPE method (blue histograms) for the pairing Hamiltonian 
for 8 particles on 8 double degenerated single-particle levels 
and $g/\Delta e=0.5$ using $n_q=4$ (left column), $n_q=6$ (middle column) and $n_q=8$ (right column) ancillary qubits.  
Shown from top to bottom are the results obtained with the Hartree-Fock (HF) (a-c), Q-PAV (d-f) and Q-VAP (g-i) states. The HF state
is the one obtained when only filling the $N/2$ lowest levels.
The vertical green and black solid lines indicate the ground state and the first excited state energies respectively. 
In the rightmost column, we show the probabilities in logarithmic scale to resolve small components in the QPE. In this case, the 
horizontal dashed lines indicate the exact eigenstates of the Hamiltonian. Note that, in each panel, the width of the histogram
corresponds to the resolution of the QPE method for a given $n_q$ (see text). }
\label{fig:appqpe}
\end{figure*}

\subsection{Quantum Phase Estimation algorithm for energy spectra} 

The use of the QPE algorithm for energy spectra is quite demanding in terms of quantum resources and is difficult to implement within
 the current NISQ period.
Nevertheless, it remains a good reference for methods that give access to both the eigenvalues and eigenstates 
of the Hamiltonian. We note that it has already been applied to the pairing Hamiltonian with a different   fermions--to--qubits encoding in Ref. 
\cite{Ovr03,Ovr07}. 
 
Here, we are interested in the eigenvalues of the Hamiltonian $H$.  As discussed in  \cite{Ovr03,Ovr07}, the constraint (\ref{eq:constphi}) to all eigenvalues is a serious limitation for the QPE application. One possible way to satisfy this constraint is to assume:
\begin{eqnarray}
V &=& \exp{\left\{-2 \pi i \left( \frac{H - E_{\rm min}}{E_{\rm max} - E_{\rm min} }\right) \right\} }, \label{eq:propqpe}
\end{eqnarray}  
where $E_{\rm min} < E_{\rm max}$ are two constants chosen so that all eigenvalues of $H$ verify  
$E_\alpha \subset [ E_{\rm min}, E_{\rm max}[$.  
As already discussed in Ref. \cite{Ovr03}, one of the drawbacks of the 
QPE is that it already requires approximate knowledge  of the eigenvalues boundaries to be applicable. 
In all the calculations presented here, we used $E_{\rm min} = 0$.  There is lots of flexibility in the choice 
of $E_{\rm max}$, the only constraint being to be above all eigenenergies of $H$. ${\cal E}_{\rm max}$ denotes below the highest eigenvalue. 
For a fixed number of ancillary qubits, the closer ${\cal E}_{\rm max}$ is to $E_{\rm max}$, the better is the precision on the eigenvalues.  
In general, the eigenvalues are unknown and the default value proposed by Qiskit is $E_{\rm max}= \sum_l |\beta_l|$  that could be inferred from Eq. (\ref{eq:betal}).   
This default value can be considered as a canonical choice and is equal  to $E_{\rm max} = \sum_p |2\varepsilon_p-g| + |g| N(N-1)/2$ for the pairing Hamiltonian. 
In the illustration below, since we have access to the true value of ${\cal E}_{\rm max}$, we simply used $E_{\rm max} = {\cal E}_{\rm max} 2^{n_q}/(2^{n_q} -1)$. 

We applied the QPE algorithm using three different initial states with an appropriate number of particles (i) The pure Hartree-Fock solution where the initial state is the Slater determinant occupying the lowest N/2 single-particle states. The corresponding energy is denoted $E_{\rm HF}$ ; (ii) The Q-PAV state obtained directly by projecting after the BCS minimization and; (iii) The Q-VAP trial state that minimized the projected energy. 
        
The QPE approach applied to the operator $V$ requires the quantum simulation of the propagator $U(\tau) = e^{-i \tau H}$ for various time intervals. Here, we follow the standard Trotter-Suzuki method \cite{Tro59,McA20} where we discretize $\tau$ into small time steps $\Delta \tau$. 
The propagator over $\Delta \tau$ is then decomposed as $U(\Delta \tau) = U_{\varepsilon}(\Delta \tau)U_{g}(\Delta \tau)$ with (for more details see \cite{Rui21}):
\begin{eqnarray}
U_{\varepsilon}(\Delta t) & = &  \prod_{p}\left(\begin{array}{cc}
1 & 0\\
0 & \exp\left(-i\left(2\varepsilon_{p}-g\right)\Delta t\right)
\end{array}\right)_p 
, \label{eq:pepsilon}
\end{eqnarray}
and for the two-body interaction part:
\begin{eqnarray}
U_{g}(\Delta t) & =& \prod_{p>q}
 \left(
 \begin{array}{cccc} 
 1& 0 & 0 & 0 \\
0 & \cos(g \Delta t) & i\sin(g \Delta t) & 0\\
0 & i\sin(g \Delta t) & \cos(g \Delta t) & 0\\
0 & 0 & 0 & 1
\end{array}\right)_{pq}.
\end{eqnarray}
In practice, we have used $\Delta \tau \Delta e \approx 10^{-2}$ that ensures good precision for the Trotter-Suzuki
method. 

We show in Fig.  \ref{fig:appqpe} the results of the QPE method with varying number of ancillary qubits $n_q=4$, $6$ and $8$ 
and for the three initial states. We see in this figure, that peaks appear rather rapidly as $n_q$ increases. As shown in the figure, 
these peaks correspond to the eigenvalues, denoted by $\{ E_\alpha \}$, of the many-body pairing Hamiltonian. We denote by $| \alpha \rangle $
the corresponding exact eigenstates and assume that the initial state decomposes as follows:
\begin{eqnarray}
| \Psi \rangle &=& \sum_\alpha c(\alpha) | \alpha \rangle.
\end{eqnarray}
For a non-degenerate state, the height of the peak corresponding to an eigenvalue $E_\alpha$ converges approximately to $| c(\alpha)|^2$ for large values 
of $n_q$. We have indeed checked that this is the case for the ground state which is well isolated from other eigenvalues in the pairing Hamiltonian. 
Another conclusion that can be drawn by comparing panels (b), (e) and (d) in Fig. \ref{fig:appqpe} 
is that the probability $p_{GS}$ of the peak corresponding to the ground state component is such that 
$p^{\rm HF}_{GS} <  p^{\rm Q-PAV}_{GS} < p^{\rm Q-VAP}_{GS}$. In other words, the overlap between the Q-VAP state and the "true" 
GS is larger than for other initial states. In panel (i), we also note 
that the Q-VAP approach has "purified" the state compared to the Q-PAV state, reducing the contributions from the excited states.       
It is interesting to mention that the QPE algorithm, once applicable on real quantum platforms will be a formidable tool to scrutinize 
the approximations that are standardly used in the nuclear many-body problem.

\subsubsection{QPE precision and convergence}

The convergence of the QPE method in a perfect quantum computer is rather well documented \cite{Nie02} and we 
give below only an illustration of this aspect together with some elements useful for the following discussions. We focus here on 
the convergence for the ground state and the first excited state and henceforth refer to $E_{\rm GS}$ and $E_{\rm 1st}$ as their energies.  

\begin{figure}
\begin{centering}
\includegraphics[scale=0.5]{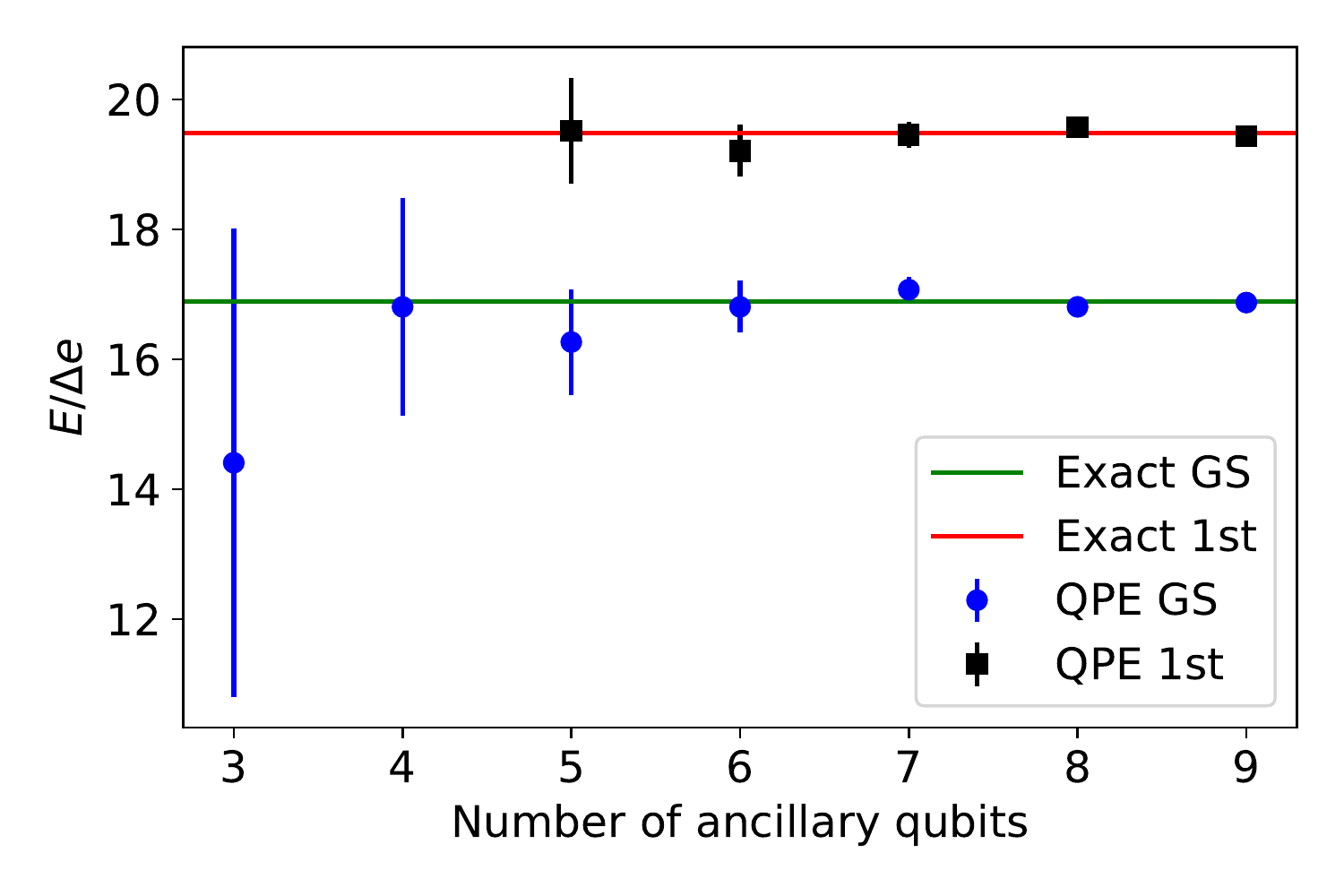} 
\caption{Illustration of the ground state (blue filled circles) and $1^{st}$ excited energies (black filled squares) obtained with the QPE method displayed in Fig. \ref{fig:appqpe} for the HF initial state. The showed energies correspond to the position of the center of the peak if present. For the 1st excited state, we do not show the energies for $n_q=3$ and $4$ since no peak can be identified. The error bars correspond to the bin size.}  
\label{fig:enerqpe}
\par\end{centering}
\end{figure}

For a given number of ancillary qubits, the accuracy of the eigenenergies in Fig. \ref{fig:appqpe} is directly illustrated by the width 
of the histogram. An analytical expression for this precision can be obtained following Ref. \cite{Nie02}. Let us follow section \ref{eq:proj} and 
assume that an eigenvalue of $V$ is written as $e^{2\pi i \varphi}$.  We then introduce the binary fraction of $\varphi$ denoted by:
\begin{eqnarray}
0. \varphi_1 \varphi_2 \cdots = \frac{\varphi_1}{2} +  \frac{\varphi_2}{2^2} + \cdots  \label{eq:binfrac}
\end{eqnarray} 
The QPE method with $n_q$ qubits gives access to the approximated values of $\varphi$ under the assumption that the binary fraction 
is truncated at order $n_q$. Denoting by $\varphi(n_q)$ the approximation, we immediately obtain an upper bound on the error for the phases:
\begin{eqnarray}
0 \le  \varphi - \varphi({n_q}) \le \frac{1}{2^{n_q}} .
\end{eqnarray}
The error on the phase can be transformed into an error on the energies. For this, we first note that
the operator $V$ can be interpreted as the propagator of the Hamiltonian $H' = (H - E_{\rm min})$ by rewriting it as 
$V = e^{-i \tau_{\rm QPE} H' }$ where we have defined: 
\begin{eqnarray}
\tau_{\rm QPE} = 2 \pi / (E_{\rm max} - E_{\rm min}). 
\end{eqnarray} 
From this, we deduce that the precision in energy is 
\begin{eqnarray}
\delta E =\frac{ \pi}{ ( 2^{n_q - 1} \tau_{\rm QPE} )} . \nonumber
\end{eqnarray}
An attractive aspect of
the QPE is that each time an ancillary qubit is added, the bin size in energy is divided by $2$. We illustrate in Fig.  \ref{fig:enerqpe} the convergence and precision for the ground state and first excited state energies. 

This attractive feature should be moderated by the fact that the number of operations to be performed in the system circuit increases substantially when a single qubit is added.     
From Fig.  \ref{fig:qpe}, we see indeed that applying the QPE algorithm is equivalent to performing a series of successive 
propagations of the system over times $2^0 \tau_{\rm QPE}$, $2^1 \tau_{\rm QPE}$, ..., $2^{n_q-1}\tau_{\rm QPE}$. The use of $n_q$ 
qubits corresponds to a propagation over a total time $\tau^{\rm QPE}_{\rm tot}$ given by:
\begin{eqnarray}
\tau^{\rm QPE}_{\rm tot}(n_q) &=& \tau_{\rm QPE} + 2 \tau_{\rm QPE} + \dots + 2^{n_q-1}\tau_{\rm QPE} \nonumber \\
&=& \left( 2^{n_q} - 1\right)\tau_{\rm QPE}. \label{eq:timeqpe}
\end{eqnarray}
This shows that every time a qubit is added in the QPE, the time evolution is essentially multiplied by a factor of $2$. 
The same scaling appears directly in the number of operations on the system circuit required to perform the QPE. 
Let us assume that we need $N_{\rm op}$ operations or gates to perform the Controlled-$V^{2^0}$ in the circuit  of Fig. \ref{fig:qpe}. This 
number of operations includes the propagation of the system up to $\tau_{\rm QPE}$ using the Trotter-Suzuki method as well as the controlled gate operations. 
Then, $2^j N_{\rm op}$ operations are required for a given $j$ to perform the controlled-$V^{2^{j}}$ gate shown in Fig. \ref{fig:qpe}.  This implies that
the total number of operations for the QPE  increases rapidly with $n_q$ and also scales like $N_{\rm tot} = \left( 2^{n_q} - 1\right) N_{\rm op}$. This scaling is extremely demanding in terms of quantum resources.   

Finally, we would like to mention that the convergence properties and, in particular, the precision are insensitive to the 
initial state. This is true for the bin size, which depends only on the $\tau_{QPE}$ and $n_q$ values in Eq. (\ref{eq:timeqpe}).  From this point of view, there is no clear advantage to using a Q-VAP state instead of the simplified state HF. We could even argue that using a simple HF state, which has a smaller initial overlap with the ground state, provides a better starting point to obtain a larger number of excited states, as can be seen in panels (c) and (i) of Fig. \ref{fig:appqpe}. The HF state also has the clear advantage of requiring far fewer quantum operations to prepare it.
We note that some new algorithms have been proposed recently to achieve faster convergence compared to QPE \cite{Cho21,Qio21} but again, we do not anticipate that these novel algorithms will benefit from an improved preparation of the initial state.  

In the following, we explore alternative methods to obtain the energy spectra with increasing accuracy, using the optimization of the initial state as in the Q-VAP technique.

\subsection{Quantum Krylov approach}

\begin{figure}
\includegraphics[width=0.7\linewidth, angle=-90]{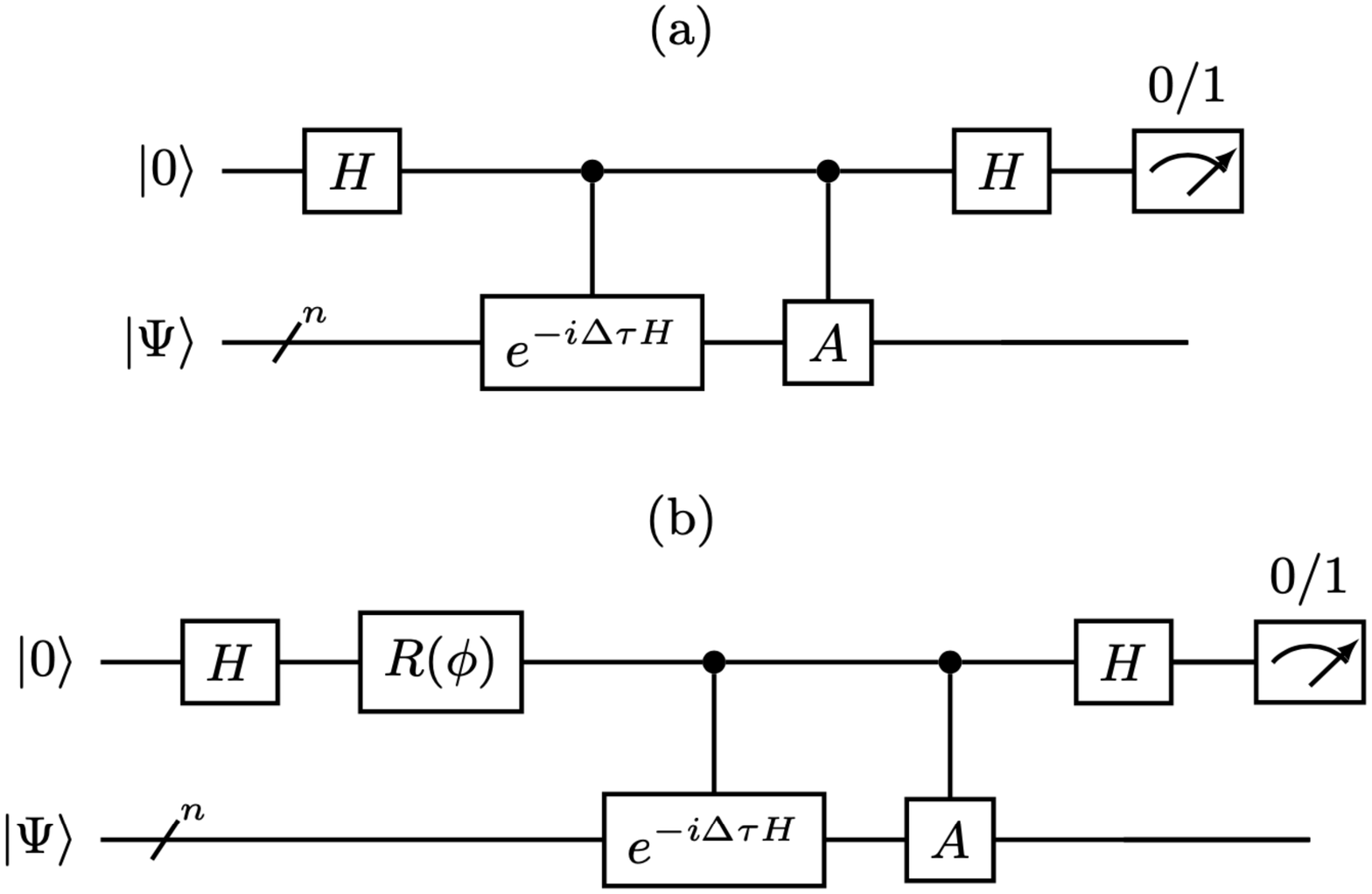}
 \caption{Illustration of the (a) Hadamard and (b) modified Hadamard test used to calculated the real and imaginary parts of the expectated value $\langle \Psi | A e^{-i \Delta \tau H} | \Psi \rangle$. 
In this circuit, $H$ is the standard Hadamard
gate while $R(\phi)$ corresponds to the phase gate where the angle
is set to $\phi=-\pi/2$. Note that in both circuits the quantity of interest is obtained from the difference $p_0 - p_1$ where $p_0$
(resp. $p_1$) is the probability of measuring $0$ (resp. $1$) in the ancillary qubit.  In the present work, we identify the operator $A$ with the 
identity for computing the overlaps (\ref{eq:oij}).  For the Hamiltonian, we consider the decomposition (\ref{eq:betal}) and compute 
the desired quantities for each operator $V_l$ separately. 
}
\label{fig:hadamreim} 
\end{figure}

A possible alternative to the QPE, is to use iterative methods that start from an initial state $| \Psi \rangle$ and gradually construct a set of states leading to a subspace of the Hilbert space where configuration interaction (CI) calculations can be achieved by diagonalizing the Hamiltonian in the reduced space. Such techniques are widely used on classical computers \cite{Saa11}. 
Among these techniques, we can mention those based on the Krylov state defined by the set of $M$ states:
\begin{eqnarray}
\{ | \Psi\rangle, ~H | \Psi\rangle, \cdots ,~ H^{M-1} | \Psi\rangle \},  \label{eq:stkryl}
\end{eqnarray} 
like the widely used Lanczos and the Arnoldi iterative methods. 
Quantum algorithms related to the Krylov space have attracted recently special attention 
\cite{Par19,Bes20,Mot20,Bes21,Sek21,Bak20,Bha20,Hau20, Bha20-b,Bha21-b,Rog20,Kow20,Rui21,Cor21,Sta20,Lau21} (see also the recent survey \cite{Aul21}). The brute force mapping of the Krylov based techniques using the reduced basis 
(\ref{eq:stkryl}) requires the precise estimates of the different expectation values $\langle H^K \rangle$ for $K \le 2M-1$. However, because the operators $H^K$ are not unitary, it is not straightforward to determine their expectation values on a quantum computer. One possible way is to obtain a similar
expression for $H^K$ as in Eq. (\ref{eq:betal}) and compute each term in the expansion separately. In this direct strategy, 
the number of terms quickly becomes very large as $K$ increases. In a recent study \cite{Rui21}, 
we explored the possibility of computing the moments of $H$ directly by successive derivatives of the generating function
$F(t) = \langle e^{-i \tau H} \rangle$. However, the precision in the estimates decreases rapidly as the order $K$ increases. 
Here, we investigate the alternative possibility of using the Quantum Krylov based methods \cite{Par19,Sta20,Cor21}.  The starting point of the approach is to replace the states (\ref{eq:stkryl}) by
the new set of states:
\begin{eqnarray}
\{ | \Psi\rangle, ~e^{-i \tau_1 H} | \Psi\rangle, \cdots ,~ e^{-i \tau_{M-1}H}  | \Psi\rangle \}. \label{eq:qkrylov}
\end{eqnarray}  
In the following, we will simply write $| \Phi_{n} \rangle \equiv e^{-i \tau_n H} | \Psi \rangle$ for $n=0, M-1$, with the convention 
that $\tau_0=0$. Our goal is to diagonalize the Hamiltonian in the reduced subspace formed by the non-orthogonal states (\ref{eq:qkrylov}). 
To this end, we introduce the overlap and Hamiltonian matrix elements:
\begin{eqnarray}
O_{ij} &=& \langle \Phi_i | \Phi_j \rangle = \langle \Psi | e^{-i (\tau_j - \tau_i ) H} | \Psi \rangle , \label{eq:oij} \\
H_{ij} &=&  \langle \Phi_i | H | \Phi_j \rangle = \langle \Psi | H e^{-i (\tau_j - \tau_i ) H} | \Psi \rangle.  \label{eq:hij}
\end{eqnarray}  
Below, for the sake of compactness, we will sometimes write $\Delta \tau_{ji} = \tau_j - \tau_i $.    
To find approximate solutions to the eigenvalue problem, we decompose the approximate eigenstates:
\begin{eqnarray}
| \xi_\alpha \rangle &=& \sum_n c_n (\alpha) | \Phi_n \rangle. \label{eq:decomp}
\end{eqnarray}  
Every eigenstate is solution of the generalized set of eigenvalue equations:
 \begin{eqnarray}
\sum_n c_n (\alpha) H_{in} &=& E_\alpha \sum_{n} c_n (\alpha) O_{in}, \label{eq:eigen}
\end{eqnarray} 
These equations correspond to standard eigenvalues equations written in a non-orthonormal basis \cite{Rin80}. 

Here we consider a hybrid quantum-classical algorithm where the computation of the Hamiltonian and overlap matrix 
elements, given by (\ref{eq:oij}) and (\ref{eq:hij}), is made on a quantum computer while the solution 
of the set of equations (\ref{eq:eigen}) is performed on a classical computer. The circuits used to compute the real and 
imaginary parts of the expectation values  (\ref{eq:oij}-\ref{eq:hij}) correspond to the standard Hadamard or modified Hadamard tests 
shown in Fig. \ref{fig:hadamreim}. 

The eigenvalue problem given by Eq. (\ref{eq:eigen}) is solved using a standard technique. In practice, starting from a set of times
$\{ \tau_i \}_{i=0,M-1}$, the various overlaps and matrix elements of $H$ are first computed using the circuits shown in Fig.  
\ref{fig:hadamreim}. This information is then transmitted to a classical computer. The eigenvalues and eigenvectors components in the reduced space are determined in two steps. First, the overlap matrix is diagonalized, resulting in a set of orthonormal states $\{ | \chi_i \rangle \}_{i=0, M-1}$ and eigenvalues $\{ \lambda_i \}_{i=0, M-1}$ for the reduced Hilbert space. Note that some of the states may not be retained for further processing if the eigenvalues are below a certain threshold $\lambda_i \le \epsilon$.  This happens when some of the states in the set $\{ | \Phi_i \rangle \}$ 
are a linear combination of the others.  After this step, the Hamiltonian is diagonalized in the basis $\{ | \chi_i \rangle \}$ leading to a set of $J \le M$ eigenvectors where $J$ is the set of states that are retained after the first step. In the following applications, we use $\epsilon = 10^{-6}$ and figures will always present results as a function of the original number of states $M$. 
\begin{figure}[htbp]
\includegraphics[width=3.5in]{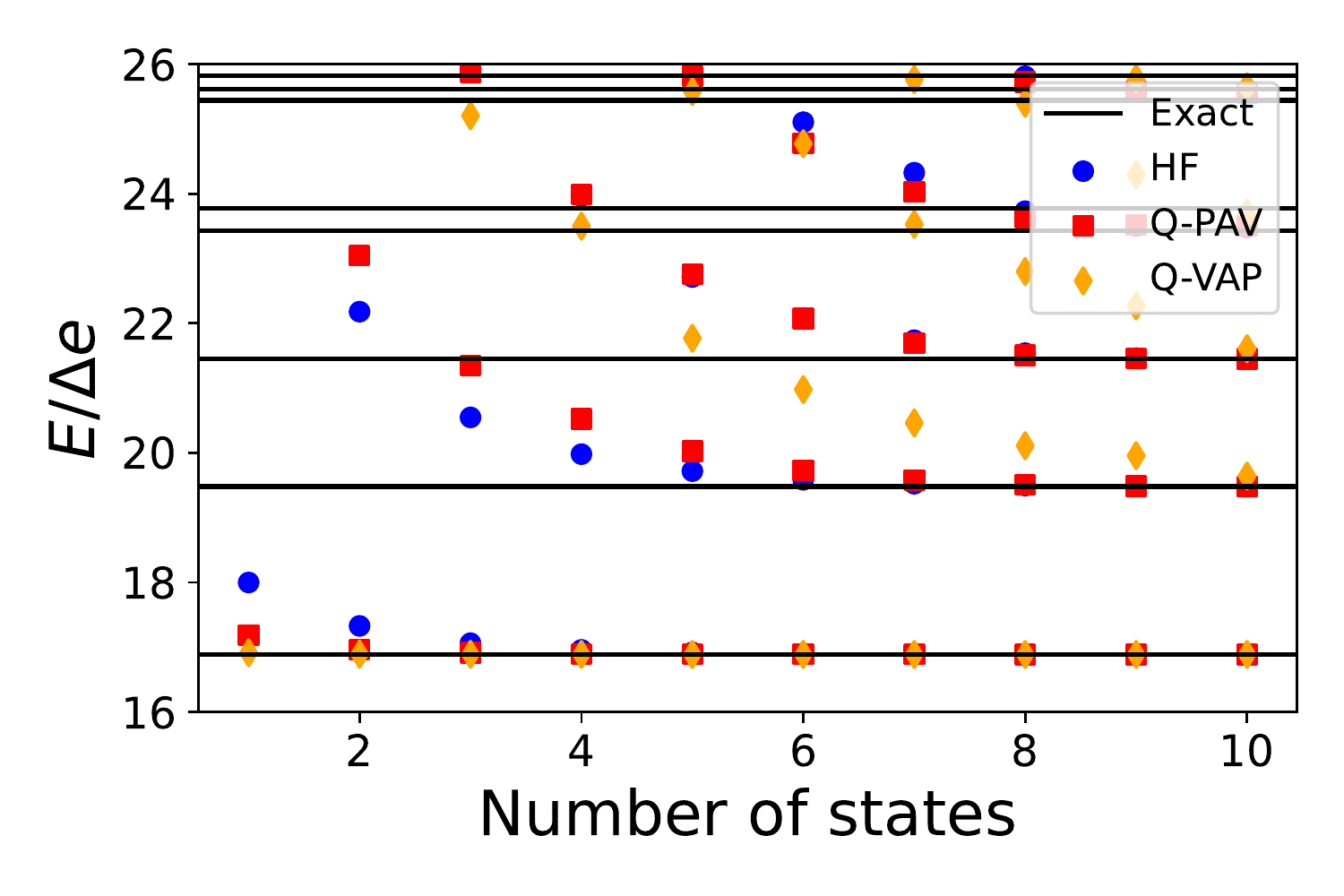}
\caption{Illustration of the energies obtained by the 
Quantum Krylov method for the pairing problem with $8$ particles on 8 levels and $g/\Delta e = 0.5$. 
The results are obtained using the set of times $\tau_i = i \Delta \tau$ for $i=0, M-1$ 
and starting from the HF (blue circles), the Q-PAV (red squares) and Q-VAP (orange diamonds).  
The approximate energies are plotted as a function of $M$. For the present figure, we used $\Delta \tau . \Delta e = 0.3$ 
 and a threshold $\epsilon = 10^{-6}$ for the rejection of states when diagonalizing the overlap matrix. 
 The horizontal black lines indicate the exact eigenenergies. }
\label{fig:espectra} 
\end{figure}

\subsubsection{Discussion on the Quantum Krylov method convergence }

Some aspects and possible improvements concerning the convergence of the quantum Krylov 
method were discussed in Ref. \cite{Cor21}. In the following, we will focus on the influence of the 
initial state optimization on the convergence of the approach by comparing the different methods to 
initialize the system. 

\begin{figure}[htbp]
\includegraphics[width=3.5in]{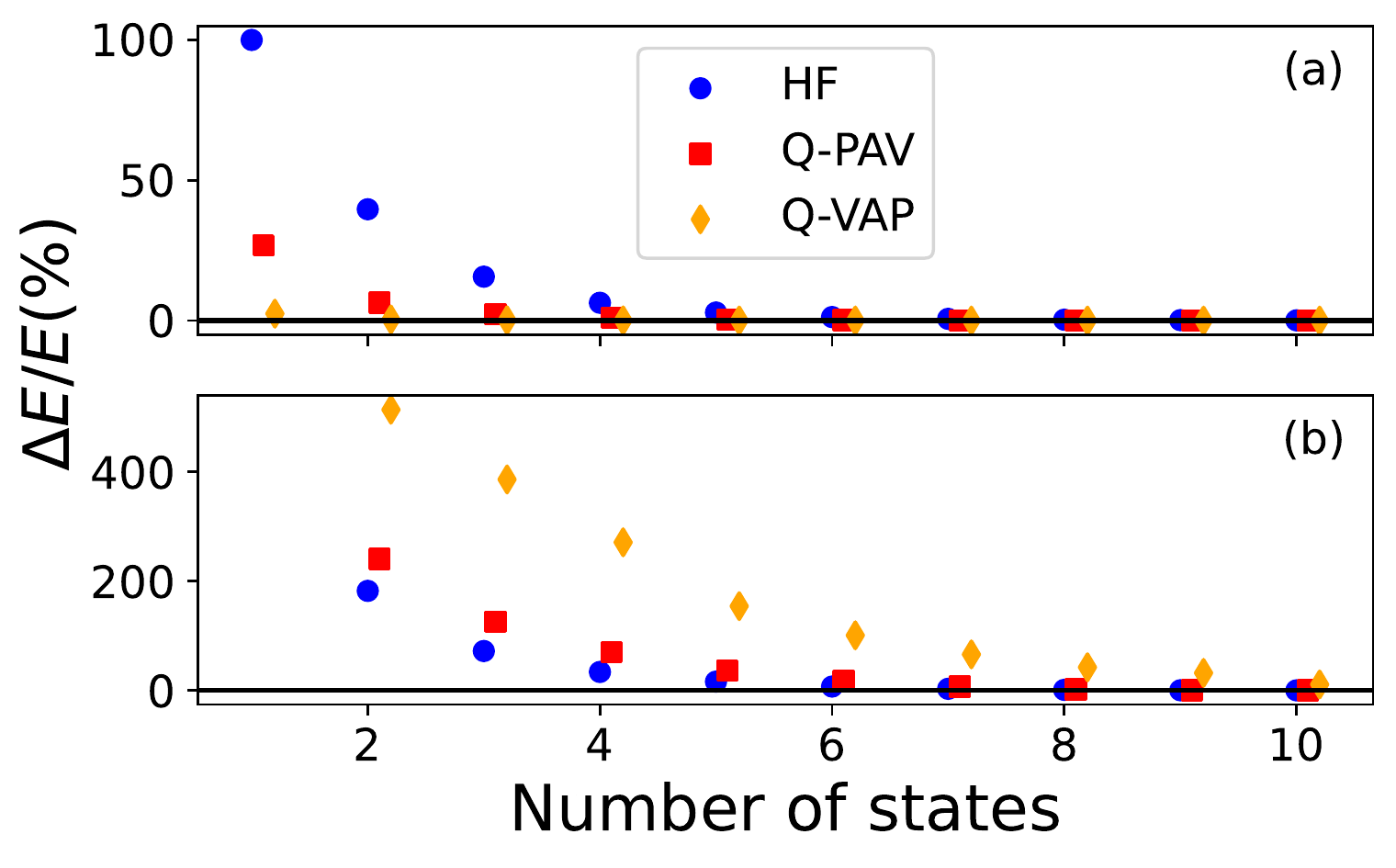}
\caption{Percentage of error defined by Eq. (\ref{eq:ecpercent}) obtained with the Quantum Krylov method for the 
ground state (a) and first excited state (b) starting from the different initial states as a function of the number of states $M$. 
The horizontal lines correspond to the exact energies. In this figure, the precision displayed for $M=1$ 
are those reported in Fig.  \ref{fig:figbcsP} for $g/\Delta e = 0.5$. }
\label{fig:focussmall} 
\end{figure}

We show in Fig. \ref{fig:espectra} the energy spectra obtained by the Quantum Krylov method starting from the HF, Q-PAV and Q-VAP states.  
In this figure, an increasing number of states $M$ is used and
the states are generated with a constant time steps $\tau_i = i \Delta \tau$ for $i=0, M-1$. 
When only one state is used, i.e., when $M=1$, the energy corresponds to the energy of the initial state. 
 We see in this figure that the energies obtained with the Quantum Krylov method converge
 towards some of the exact eigenvalues regardless of the initial conditions. The rapidity of convergence clearly depends on the specific targeted energy and 
 on the initialization procedure. To illustrate this aspect, we focus in Fig. \ref{fig:focussmall} on the accuracy of the energy obtained for the ground state and the first excited state using the percentage of error defined in Eq. (\ref{eq:ecpercent}). 
For the ground state shown in panel (a), the convergence for the Q-VAP initial state is much faster, showing 
the net advantage of using the optimization at the level of the symmetry projected state. 
The rapid convergence observed for this initial state can be directly attributed to the strong overlap between this initial state and the ground state, as also shown in panel (g-i) of  Fig. \ref{fig:appqpe}. The advantage of the Q-VAP initial state clearly breaks down for the first 
excited state. In this case, a simple HF initialization is able to achieve the best convergence. As shown in Fig. \ref{fig:espectra}, the same conclusion holds for all excited states. Indeed, in this figure, we see that the convergence towards excited states is similar for the HF and Q-PAV state and in all cases faster than for the Q-VAP initialization. We finally mention that some excited states are not obtained in Fig. \ref{fig:espectra}  because they are either not present in the initial state, or their components
are initially below the threshold $\epsilon$, or because the size of the Quantum Krylov basis is not large enough.  

\subsubsection{Comparison between QPE and Quantum Krylov}

\begin{figure}
\begin{centering}
\includegraphics[scale=0.5]{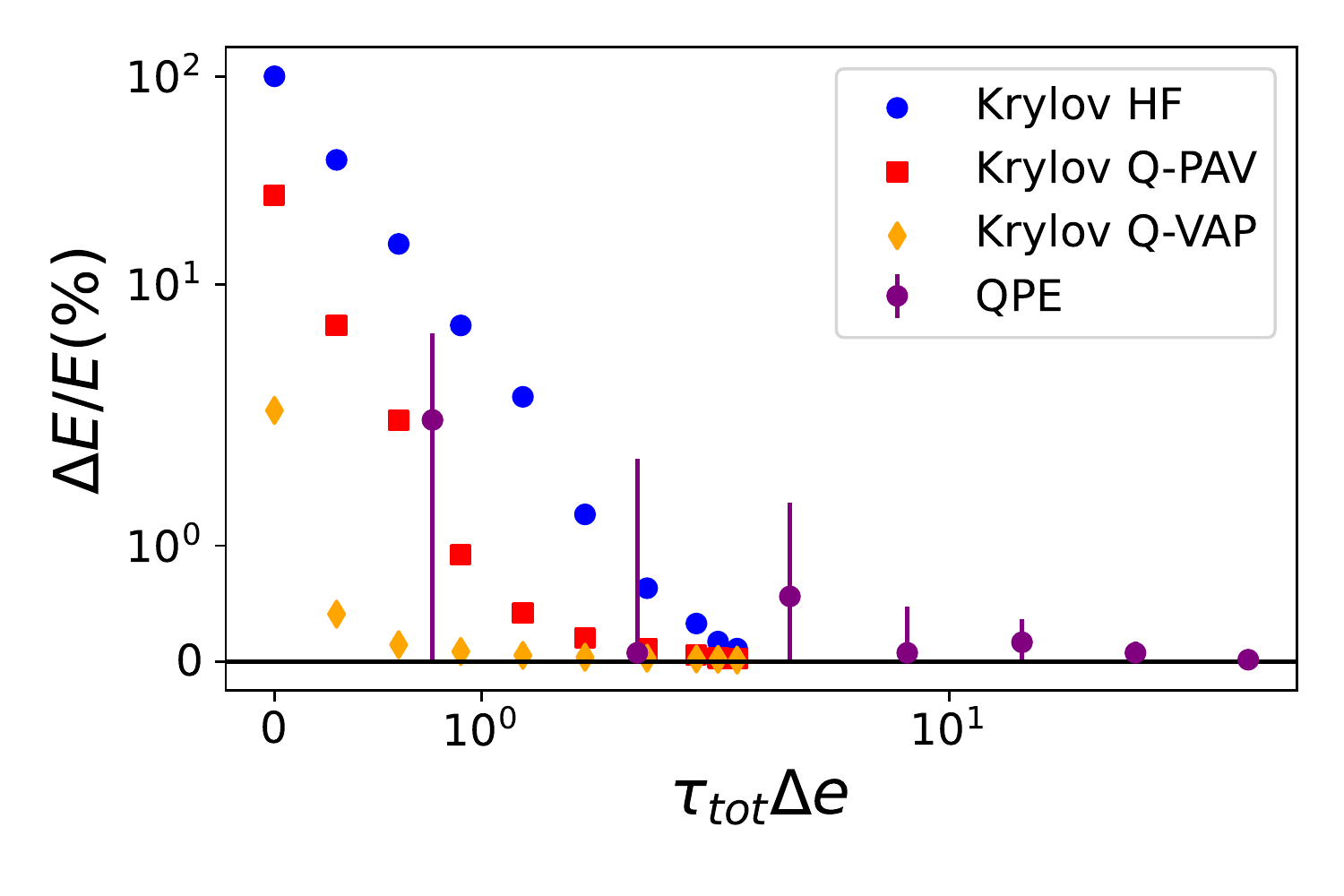}
\caption{Illustration of the time evolution $\tau_{\rm tot}$ 
needed to obtain a certain precision for the ground state energy using the Quantum Krylov method for different initial states
(HF, Q-PAV and Q-VAP). These times are compared to the time (\ref{eq:timeqpe}) necessary for the QPE method for $n_q= 3, 4, \dots, 9$ (purple filled circles). The errorbars shown for the QPE case are computed from the bin size of Fig. \ref{fig:appqpe}.
}  \label{fig:prectau}
\par\end{centering}
\end{figure}
We discuss here some aspects of the two methods used for post-processing (QPE vs Quantum Krylov). 
First, we note that the two post-processing strategies are different in nature since the QPE is purely quantum-based while the Quantum Krylov 
method falls into the class of hybrid quantum-classical computations. Moreover, the results of the methods 
are also slightly different. In the absence of noise and assuming that the number of qubits that can be used is unlimited, 
the QPE approach gives a priori access to the eigenstates and eigenvalues with arbitrary precision in the full Fock space. 
The Quantum Krylov method gives approximate eigenvalues and components of the eigenstates in a reduced subspace of the total 
Hilbert space. This difference in outcomes should be kept in mind when comparing the two methods as we do below.     

A first evident advantage in favor of the Quantum Krylov method is the circuit length. 
The Quantum Krylov only requires one extra ancillary qubits compared to the QPE for which the number of extra qubits needed varies with the desired accuracy. Another aspect is the number of 
operations itself to reach this accuracy. Since both methods require the controlled-$U$ operation with $U(t) = e^{-i t H}$,
and since both are implemented here using the Trotter-Suzuki method, a compact way to compare the number of operations is to compare the time 
${\rm \tau}_{tot}$ over which the system should be evolved to reach a certain precision on the energy. This time is given by Eq. (\ref{eq:timeqpe}) for the QPE case. In this case, we have shown that the precision is rather independent of the initial state. 
For the Quantum Krylov method, the precision achieved for instance on the ground state, 
depends on the optimization and not on the initial state (see Fig. \ref{fig:focussmall}). For the Quantum Krylov method, the maximum total 
time needed is identified as the maximal value in the set of times $\{ \tau_i \}_{i=0, M-1}$. If we assume that the times 
are sorted in ascending order, we have simply $\tau^{QK}_{\rm tot} = \tau_{M-1}$. 

In Fig. \ref{fig:prectau} we compare the precision on the ground state energy as a function of the total time of propagation 
in the two methods. Regardless of the initial condition, we see that the simulation time required to achieve a certain precision for the ground 
state energy is at least an order of magnitude smaller  for the Quantum Krylov method compared to the QPE approach. 

\subsubsection{Improving the convergence for excited states}

The short simulation time required for the Quantum Krylov approach is a major advantage over the QPE method. 
This conclusion also holds for the first few lowest excited states shown in Fig. \ref{fig:espectra}, regardless of the initial state. 
Nevertheless, this figure shows that the most optimized initial state, i.e., the Q-VAP ground state, which has the fastest convergence to the true ground state, is the least effective for the excited 
states. Even the crude HF approximation leads to fastest convergence in the latter case. 
Such feature probably stems from the strong purification of the Q-VAP ground state that leads to very small overlaps of the projected state
with the exact excited states as shown in Fig. \ref{fig:appqpe}-i. 
\begin{figure}[htbp]
\includegraphics[width=3.2in]{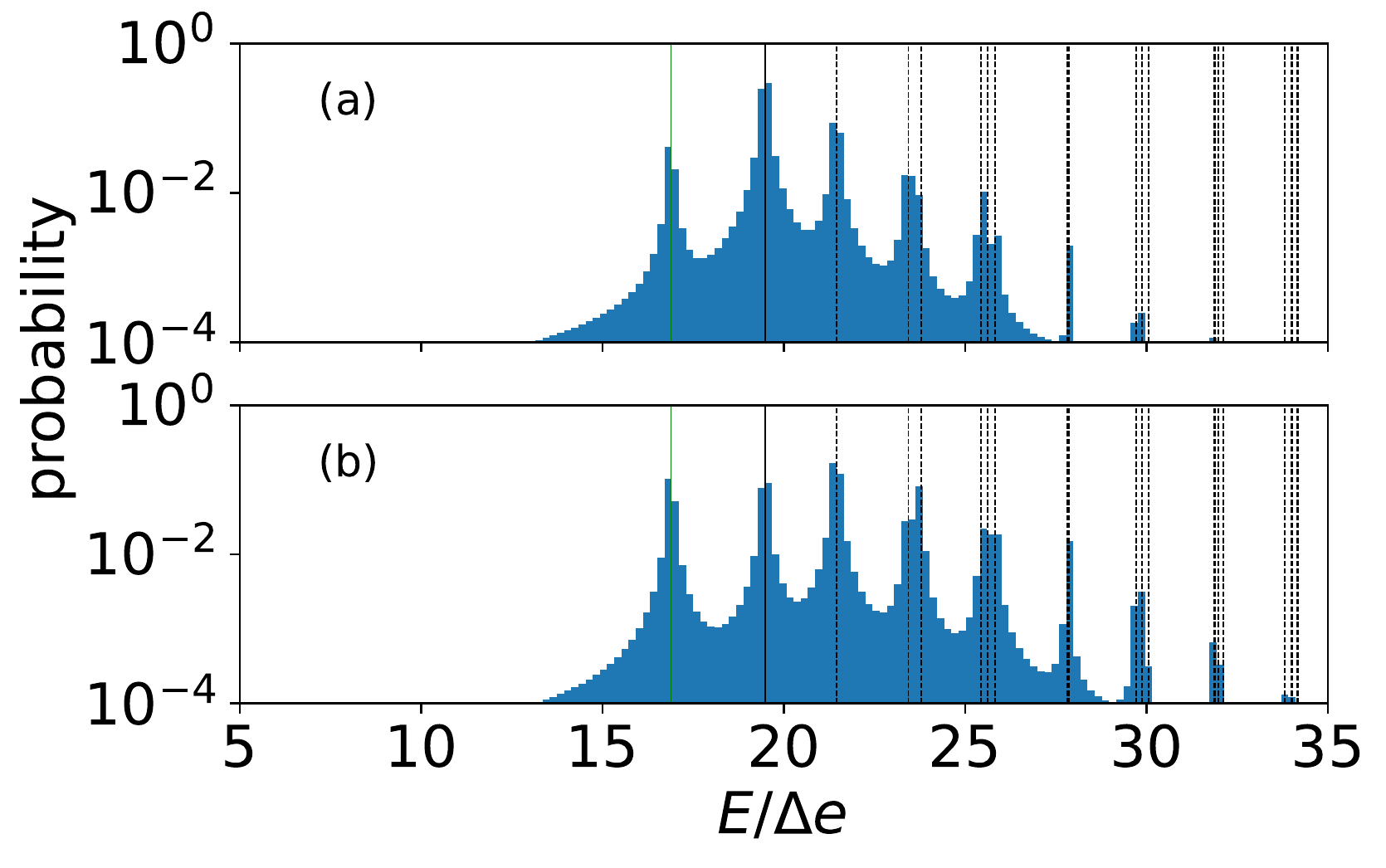}
\caption{Same as panel (i) of Fig. \ref{fig:appqpe} where the QPE is applied with $n_q=8$ ancillary qubits but using an initial 2QP or 4QP state projected onto a given number of particles. 
The illustration is performed for $8$ particles on $8$ levels and $g/\Delta e=0.5$. The initial states used in the QPE correspond to states given by Eq. (\ref{eq:bcsexcited})
projected onto $A=8$. Panel (a) uses a 2QP state where the QP are those associated to the $3^{rd}$ single-particle level. Panel (b) is associated to a 4QP state 
with QPs states associated to the $3^{rd}$ and $4^{th}$ single-particle states. }
 \label{fig:qpeexcited}
\end{figure}

One can take advantage of our knowledged of  the BCS theory to improve the Q-VAP convergence. In the BCS framework, starting from the ground state 
(\ref{eq:bcsansatz}), excited states are generated by quasiparticle (QP) excitations. In the specific case we consider, where we assume no pair breaking, i.e., zero seniority,
the excited states correspond to $2$QP, $4$QP, $\ldots$ excitations. Starting from the state  (\ref{eq:bcsansatz}), a $2k$QP excitation takes the form:
\begin{eqnarray}
| \Psi_{i_1, \cdots, i_k } (\{ \theta_p \}) \rangle =&& \bigotimes_{m=1}^{k}  \left[ - \cos(\theta_{i_m} | 0_{i_m} \rangle + \sin(\theta_{i_m}) |1_{i_m} \rangle \right]  \nonumber \\
&&  \hspace*{-0.5cm}
 \bigotimes_{p \ne (i_1, \cdots, i_k)}\left[ \sin(\theta_p) | 0_p \rangle + \cos(\theta_p) |1_p \rangle \right]. \label{eq:bcsexcited}
\end{eqnarray}      
These states are associated to a mean-field energy given by:
\begin{eqnarray}
E_{i_1, \cdots , i_{k}} &=& {\cal E}_0 + 2 \sum_{m=1,k} {\cal E}_{i_m}
\end{eqnarray} 
where ${\cal E}_0$ is the BCS energy while ${\cal E}_i$ is the quasiparticle energy. The latter energy is given in the present model by
${\cal E}_i = \sqrt{(\varepsilon_i - \lambda)^2 + \Delta^2}$ where $\lambda$ is the Fermi energy and $\Delta$ is the pairing gap. At the mean-field 
level, all excited states given by Eq. (\ref{eq:bcsexcited}) are orthogonal to the BCS ground state (\ref{eq:bcsansatz}). 
Moreover, we see that the lowest excited states are obtained by 2QP excitations associated to the single-particle levels that are close to the Fermi energy.

To improve the convergence for the excited state in the Q-VAP approach, we tested the possibility of replacing the Q-VAP ground state 
in the post-processing by one of the $2k$QP states given by Eq. (\ref{eq:bcsexcited}). More precisely, we proceed as follows (i) The Q-VAP ground state 
is found using the variational optimization discussed earlier; (ii) After this optimization, we construct one of the states given by (\ref{eq:bcsexcited}) 
without changing the values of $\{\theta_p\}$ obtained in step (i). Note that if we used directly the state (\ref{eq:bcsexcited}) in the optimization, one would converge 
to the Q-VAP ground state since the QP excitation can be identified with the original ansatz   (\ref{eq:bcsansatz}) provided that $\theta_{i_k} \rightarrow \theta_{i_k}+ \pi/2$. 
Due to the last relation, the same circuits can be used to construct the symmetry-breaking $2k$QP excited states by shifting some of the angles accordingly; (iii) the SB excited 
state is then projected onto a given particle number and used for further post-processing (QPE or Quantum Krylov). 

The orthogonality between the QP states and the BCS ground state is not preserved after projection. Nevertheless, one might expect the SR state constructed from QP excitations to have smaller overlap with the true ground state, while the contributions of the true excited states increase compared to the original Q-VAP vacuum. The QPE approach turns out to be a very useful tool to confirm this and to analyze the projected QP excited states. We show in Fig. \ref{fig:qpeexcited} the results of the 
QPE approach starting from such states with 2QP and 4QP excitations. In this figure, we clearly see the increase in the excited state components for the multiple QP excitations.  

\begin{figure}[htbp]
\includegraphics[width=3.5in]{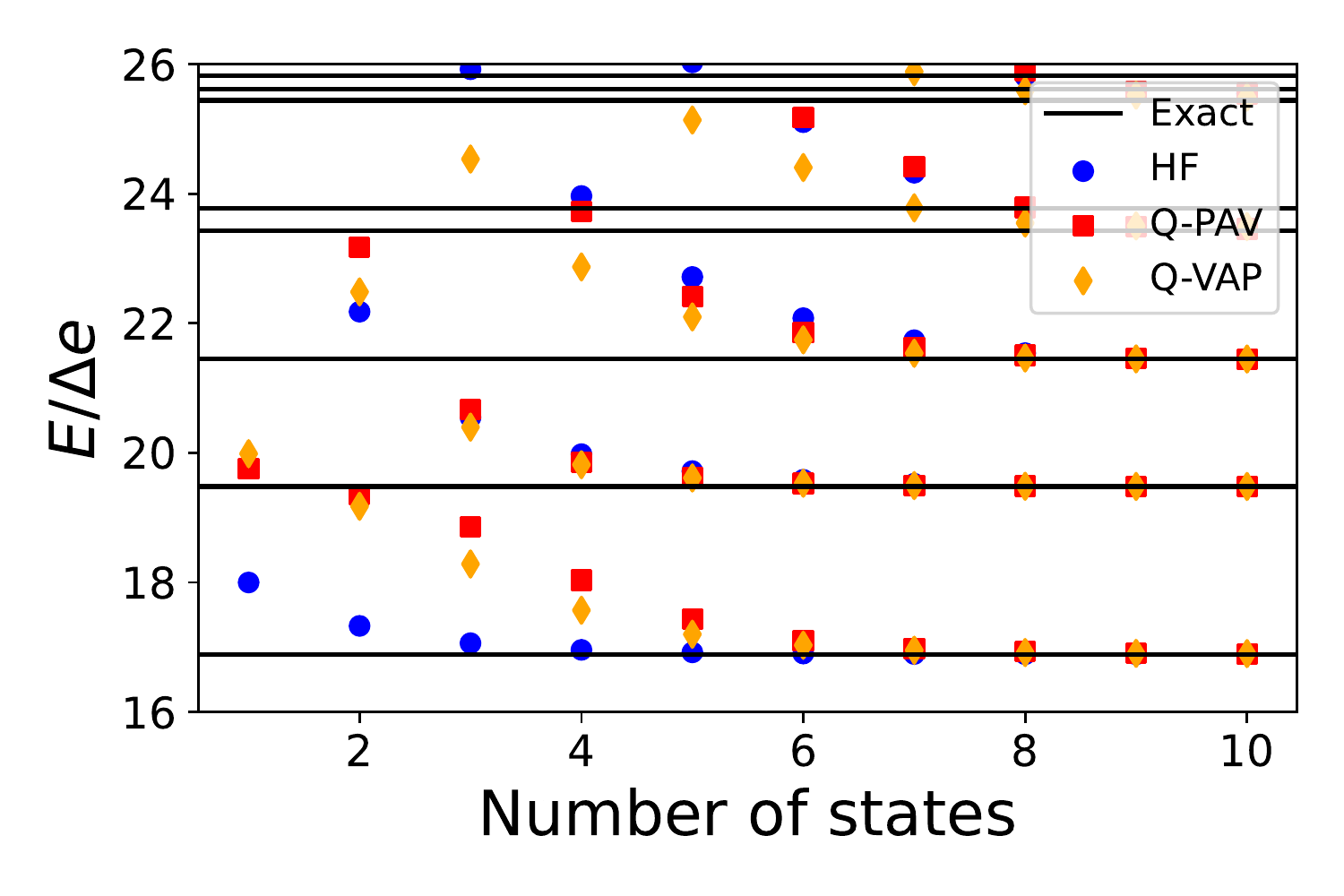}
\caption{Same as figure \ref{fig:espectra} starting from an excited state used in panel (a) of Fig. \ref{fig:qpeexcited}.}
\label{fig:espectra_1} 
\end{figure}

An illustration of the results obtained with the Quantum Krylov approach is shown in Fig. \ref{fig:espectra_1} for the initial projected 2QP state used in panel (a) of Fig. \ref{fig:qpeexcited}. 
We clearly see two consequences of using the modified initial states on the convergence. First, the use of 2QP excited state instead of the ground state clearly worsens
the convergence towards the ground state.  However, in parallel, we also observe a clear improvement of the convergence towards the first low-lying excited states.     
For the first excited state, the convergence towards excited states is significantly improved compared to the Q-VAP 
results shown in Fig.  \ref{fig:espectra}. For this state, the convergence is comparable to the HF case. A careful 
analysis shows that it is even slightly better in the Q-VAP case. For higher energy states, we clearly see that the convergence 
is strongly improved in the Q-VAP case and, in all cases it outperforms the HF or Q-PAV results. We have performed systematic studies 
by changing the quasiparticles that are used for the excitation or by performing increasing number of QP excitations. We have always improved the convergence of the quantum Krylov compared to the case without QP excitations. However, we should mention that the convergence speed depends on the type of excitation and in general it is quite difficult to predict the improvement a priori.  Nevertheless, the result shown in Fig. \ref{fig:espectra_1} is encouraging for future applications.

\section{Conclusion}     

In the present work, we  first discuss how the standard strategy consisting in breaking symmetries and restoring them on a quantum computer
can be formulated and combined with quantum variational methods. This strategy leads to highly entangled many-body states,
 which are often very efficient to describe many-body quantum systems with spontaneous symmetry breaking.  
We show that these states can be used as optimized 
initial states for further processing on a quantum computer. Such processing is illustrated here using the QPE method 
and the Quantum Krylov approach. Both techniques prove to be very efficient in obtaining the ground state energy 
when the initial state is the Q-VAP ground state. However, the advantage of using this state compared to a crude symmetry preserving 
HF approximation is lost
 when determining the excited state energy. We show here that of use of projected QP excited state can significantly 
improve the convergence towards the excited states energies.
   
The use of projection in the variational method is quite demanding in terms of quantum resources. When used prior to the QPE, we find that the 
Q-VAP approach strongly purifies the projected state towards the ground state. An initial trial state that is very close to the exact ground state 
is not necessarily an advantage in itself, especially if one wants to gain insight into the excited states, as shown in Fig. \ref{fig:appqpe}.
We show here that the situation is different for the Quantum Krylov technique. In this case, an optimized trial state significantly improves the convergence and consequently reduces the quantum resources needed for the propagation of the systems. In this case, a clear advantage is observed in the use of projected optimized states.

\section*{Acknowledgments}

This project has received financial support from the CNRS through
the 80Prime program and is part of the QC2I project. We acknowledge
the use of IBM Q cloud as well as use of the Qiskit software package
\cite{qiskit} for performing the quantum simulations.

\end{document}